# De-Virtualizing Social Events: Understanding the Gap between Online and Offline Participation for Event Invitations

Ai-Ju Huang[1], Hao-Chuan Wang[2], Chien Wen Yuan[3]

[1,2]Institute of Information Systems and Applications
& [2]Department of Computer Science
National Tsing Hua University, Taiwan
[1]s100065506@m100.nthu.edu.tw,
[2]haochuan@cs.nthu.edu.tw

[3]Department of Communication
Cornell University
Ithaca NY 14850 USA
cy294@cornell.edu

**ABSTRACT**
One growing use of computer-based communication media is for gathering people to initiate or sustain social events. Although the use of computer-mediated communication and social network sites such as Facebook for event promotion is becoming popular, online participation in an event does not always translate to offline attendance. In this paper, we report on an interview study of 31 participants that examines how people handle online event invitations and what influences their online and offline participation. The results show that people's event participation is shaped by their social perceptions of the event's nature (e.g., public or private), their relationships to others (e.g., the strength of their connections to other invitees), and the medium used to communicate event information (e.g., targeted invitation via email or spam communication via Facebook event page). By exploring how people decide whether to participate online or offline, the results illuminate the sophisticated nature of the mechanisms that affect participation and have design implications that can bridge virtual and real attendance.

**Author Keywords**
Participation; social network sites; computer-mediated communication; online events.

**ACM Classification Keywords**
H.5.3 **[Group and Organization Interfaces]:** Computer-supported cooperative work. H.5.m **[Information interfaces and presentation (e.g., HCI)]**: Miscellaneous.

**INTRODUCTION**
Social network sites (SNSs) are increasingly used to support event planning, such as recruiting and mobilizing people around the world [17, 26]. The use of SNSs has many benefits, including reaching out to potential event participants and collecting pre-commitment to attending an event through online social networks [17]. This approach facilitates resource allocation in terms of event planning and personnel deployment.

However, although the cost of initiating and managing events online tends to be low, event participation outcomes may not be as promising. One of the downsides to use SNSs or other computer-mediated communication (CMC) tools to disseminate and manage event invitations is that one's online response to an event page or a web forum does not always translate to actual attendance offline [12, 15, 25]. Anecdotal evidence often shows that anxious hosts continually seek ways to solve the problem of people hitting "Join" or "RSVP" via SNSs (e.g., Facebook) or other e-invitation sites but ending up with no show.

We broadly define event participation as any form of response to online (i.e., over technological mediation) or offline (i.e., face-to-face) event invitations. A gap may exist between online and offline participation. Using an online event call to motivate offline participation may be difficult because the cost of offline participation may be high [35], and the properties of certain existing technological designs may not be effective in motivating offline participation.

Research around the issue has focused on participation in the context of political and social movements (e.g., demonstration, election etc.). Previous studies have investigated how political events can be successfully mobilized via Facebook [16], how different media (e.g., technologically-mediated versus face-to-face) affect participation in online and offline activities [26, 35], and how messages exchanged through online social networks affect the behavioral outcomes [3]. However, few studies attempt to take a broader perspective to include non-political, everyday, and personal events such as birthday parties or cultural activities into consideration. Little is known about how people make decisions about online invitations in general.

From the design perspective, this possible mismatch between response choices on SNSs or other CMC tools and people's offline behaviors can be a signal of design failure. Online response choices are not necessarily good indicators of offline behaviors, which raises concerns about whether the original designs, such as features for event planning on SNSs, meet the intended design goals.

As event planning of all types is increasingly virtualized through the use of communication technologies, it is important to examine how event invitations move from the offline, socio-physical world to the online space and are disseminated through computer networking. Moreover, it is valuable to tackle this issue through users' perspective by exploring how they perceive and react to this change. To guide our investigation, we conceptualize the challenge of motivating offline behavior through online communication as a bottleneck in the *de-virtualization of events*. Not all events require offline participation, and a lack of offline participation may not be a problem for some events. However, in circumstances in which offline participation is required, understanding the gap between online and offline participation and the best way to de-virtualize these events is crucial. Without successful de-virtualization, it may be difficult to ground event planning and resource allocation in practice. Moreover, repercussions may follow, such as deteriorated social relationships between hosts and invitees as well as users' perceptions of the functions of the interfaces, without successful de-virtualization.

This paper contributes to an understanding of how people handle event invitations, ranging from personal events to social movements, that require offline participation when the initiation, development, communication, and some forms of participation in the events are virtualized and occur in the online space. We present the results of interviews in which we examined people's management of event invitations as well as their concerns and reasoning. We asked interviewees to recall their previous experiences in handling event invitations. We investigated whether gaps existed between online and offline forms of participation. Finally, if gaps existed, we examined the social mechanisms behind these gaps and the implications for future design.

## BACKGROUND

### (De-)Virtualization of Social Events

We observe a trend of social activity virtualization from the socio-physical, offline world to computer-mediated, online spaces. This trend is not surprising given that the costs of offline participation, such as the physical planning and attendance of an event, can be much higher than the online equivalents, such as chatting through instant messages or posting on a forum. Offline participation tends to be financially more expensive and more time consuming than many forms of online participation. In the work domain, it has become common to replace some face-to-face meetings with online discussion to reduce costs [19, 40]. Some collaboration may even occur only "virtually" through the mediation of technology, such as Wikipedia or human computation games that require a large crowd of people around the world to interact and contribute [36].

SNSs provide features that support the initiation and management of social events [16, 17]. Facebook's event pages, for example, allow users to create social events, send event invitations to others, and gather people's support and attendance decisions. For hosts, this is a convenient way to spread information about an event and reach a group of guests, and this approach requires less preparation time than other methods. It is easy for a guest to RSVP to an event to which s/he is invited through Facebook event pages. Event-supporting technologies of this sort are designed to support social event planning [21], although the actual effects of existing designs remain unclear.

Note that not all social events can be virtualized. There are cases in which physical presence and attendance are instrumental to the success of the event, which can be as personal as a birthday party or as public as a political demonstration. For example, researchers have begun to observe a phenomenon called slacktivism (combining "slacker" and "activism"). It is used to describe the scenario in which people engage in low-cost and low-risk action online without participating in related in-person activities [30]. To event organizers, online participation, in the form of such actions as electronic petitions or information sharing, is typically considered a mediating step preceding offline participation; presence at actual social events is still expected [23, 25]. However, some participants may behave differently, treating online participation as the end rather than the means to an activity. Most research now focuses on the gap between online and offline participation for political or public events, and work on participation in everyday, personal events remains limited.

People's restriction of their event participation to the online space calls into question the impact of such online participation. The lack of a real-world presence does not necessarily indicate a lack of impact. A recent study shows that people who have signed an online petition are more likely to donate to charity than those who have not [23]. However, when offline participation is vital to the success of an event, whether personal or public, an understanding of what factors influence offline presence can better reveal the properties of event virtualization.

### Social Connection, Interaction, and Participation

Recent work shows that social ties and social interaction contribute positively to event participation. For example, people are more willing to accept invitations sent by people they already know [28]. Zhang et al. study the integration of online and offline social interaction, noting that people's role identity in sport clubs on Facebook shapes their degree of participation [41]. This work highlights the individual's identity as a key factor in calls for offline action. Bond et al. explore how messages exchanged online and the strength of social ties affect political mobilization. Their results reveal that information distributed through multiplex networks in which people are connected both online and offline trigger event participation [3].

Along these lines, there is reason to believe that SNSs and online communication may encourage offline participation through the moderation of social relationships. Prior studies have made similar observations, showing that some social activities appear to benefit from the use of CMC and SNSs [16, 31]. The interactive features of SNSs (e.g., Facebook event pages) are designed to support information exchange and socialization online. Thus, it is possible for people to strengthen social connections when they socialize with their online contacts increasingly more over time [2, 10, 37]. In addition, online social interaction may facilitate the building of common ground [7], which is instrumental for collaboration among people. Therefore, technology-mediated interaction may positively contribute to offline participation decisions through increased interactivity and social connectivity.

In summary, although real-world social ties and interaction have been established as crucial to offline participation, the interactive features of SNSs may compensate for the limitations of virtualized, online communication, leading to a large amount of offline participation.

**Medium Difference and Specificity**
Different communication media have different properties or features and influence people's communication behaviors accordingly (cf. [38]). For example, in sense-making communication, rich, visible media (e.g., face-to-face or video) can often lead to more efficient communication because the establishment of mutual knowledge or common ground is facilitated when visibility is afforded [34].

Because of different media affordances, issues concerning cross-media event planning, such as the processes of the virtualization (from offline to online) and de-virtualization (from online to offline) of social events, may benefit from focused discussion on the role of media characteristics in event planning. SNSs such as Facebook allow people to express intentions to participate online by clicking "Like" or "Join" on an event page, for example. However, it is unclear whether symbolic support of this sort is representative of offline behaviors. One possibility is that people may perceive it to be more suitable to take certain types of action online and other types offline. Initial studies show that certain types of online communication (e.g., emails) tend to trigger only online behaviors, whereas face-to-face communication tends to better motivate offline behaviors [35]. In a similar fashion, SNSs may be more useful in calling for virtual support (e.g., "join an online group"), although perhaps not in-person attendance.

Another characteristic of SNSs is that they may become public or semi-public spaces (e.g., Facebook walls) where social interactions between people are displayed to irrelevant third parties [4]. Information provided on many types on SNSs, including personal or interactional information, allows users to manage their representations of themselves [5]. Following Goffman's theory of impression management [14], we can view SNSs as a front stage on which individuals present their ideal selves [4, 18]. When displayed publicly to SNS friends, reactions to event invitations can be part of one's online (i.e., onstage) self-presentation and are distinguishable from offline images and behaviors. It is possible for people to join groups on Facebook without any interest in offline participation. Participation in online groups or an agreement to attend an event may simply be a way to maintain one's personal profile [9] or present one's views or interests [15].

**THE CURRENT STUDY**
Previous work has revealed a complex set of conjectures and observations regarding the relation between online communication and participation through the use of any form of CMC and SNS and offline participation in events. To resolve this issue, it is important to study individuals' perceptions and strategies for handling technology-mediated event invitations.

We conducted an in-depth interview study that explored how people handle event invitations distributed via communication technologies and what factors influence people's online and offline participation behaviors.

**METHOD**
**Participants**
We recruited our interviewees from the largest Bulletin Board System (BBS) in Taiwan, *ptt.cc*. According to official ptt.cc information, the site has a total of approximately 1.5 million users [29]. To obtain a sample of the general public, we posted the recruiting message on a local discussion board (Hsinchu) and a board specific to the recruitment of questionnaire respondents and study participants (Q_ary). The recruiting message contained the research description and a link to a Google Document form to sign up for the study.

A total of 31 Taiwanese interviewees participated in the study (17 male, 14 female). Of the participants, 18 were college students, and 13 were non-students with various occupations (e.g., software engineers, salespersons, preschool teachers). In terms of age distribution, all were under 45. Fifteen interviewees were between 18 and 25 years old, 12 were between 26 and 35, and the remaining four were between 36 and 45. To address our research question, our interviewees were required to be CMC or SNS users and to have experience in receiving and handling event invitations mediated by some type of communication technology.

**Materials and Procedure**
This study identified people's experiences in handling technology-mediated event invitations. Before the interviews, we created an online survey using Google Documents that asked interviewees to recall and provide information about events they had attended or agreed to attend but ultimately did not attend. To allow the interviewees to provide information on the events documented during the interview, the survey asked for the title, the hyperlink (if available), and the medium used to mediate the event invitation (e.g., email, instant messaging, Twitter, Facebook, Google+, BBS). At the end of the survey, we also collected demographic information on the respondents.

One week later, we conducted one-on-one interviews and asked our interviewees to reflect on their experiences handling event invitations with the aid of information collected using the pre-interview survey. The semi-structured in-depth interview process allowed the interviewees to articulate how they grappled with the issues addressed in our research. We asked them what events they reported in the survey, how they handled the event invitations, and whether they actually attended the events. To support and better customize each interview, we checked the event hyperlinks our interviewees provided prior to the interview whenever necessary.

Each interview lasted approximately 40 minutes, and all of the interviews were audio-recorded. Most of the interviews were conducted face-to-face at the university, with the exception of two that were conducted over Skype because the interviewees could not meet in person. The audio records were transcribed verbatim, including interviewees' emotional reactions (e.g., laughing) and moments of hesitation. All of the interviews were conducted, transcribed, and analyzed in Mandarin Chinese.

**Data Analysis**
To determine whether a gap exists between people's online and offline participation, the size of the gap, and why such a gap may exist, we first identified the event types recalled by our interviewees. Then, we examined how the interviewees' responses to event invitations depended on the communication media, the interface design, and their relationships and social interactions with the event host and other invitees. Important themes were identified and are presented in the following section. The transcripts were analyzed in an iterative manner until salient themes emerged. All quotes presented in the next section were translated into English for the sake of clarity in the paper.

*Event Classification*
In the transcripts, only specific referenced events were counted and analyzed. For example, if an interviewee expressed a specific experience by making a statement such as, "Once, there was a high school reunion…," referring to a particular time or place, we identified such an incident as one event. Conversely, if a description such as "I always ignore invitations to demonstrations" was given, we did not identify the statement as referring to any type of event because it refers to a general experience rather than a specific case.

*Types of Responses*
There are four possible ways to handle online event invitations. The first is to accept the invitation by, for example, leaving a message expressing an intention to attend by clicking on the "Join" button on the Facebook event page. The second is to decline the invitation by conveying one's unavailability, such as by clicking on the "No" button on Google+. The third is to provide an uncertain response, such as clicking on the "Maybe" button on a Facebook event page. The last is to ignore the invitation by not responding. In the transcript, we identified the respondents' offline behavior for each event: whether the person attended the event (i.e., attendance) or did not attend (i.e., absence).

*Emerging Themes*
During the coding process, we highlighted recurring ideas that represented people's perceptions, thoughts, and behaviors with regard to the online event invitations across interviews. Important themes are identified and presented in this paper.

**RESULTS**
In this section, we first present the results of the event classification. We determined the frequency of different types of online and offline event participation to establish whether a gap exists between online and offline participation. We then examined the themes that emerged from the coded transcripts in an attempt to explore the reasons behind people's decisions and behaviors.

First, we confirmed that a gap exists between online and offline participation. We discovered that the communication platform and the platform's technical features matter when people evaluate events and make decisions accordingly. We then identified themes that address how people perform evaluations and make decisions. The themes revolve around the nature of the events, including public and private events, and people's relationships with the event hosts.

**Gap between Online and Offline Participation**
Based on events identified and extracted from the transcripts, we determined whether online responses were consistent with offline behavior. In total, 79 events were reported during the pre-interview survey and the interviews. The majority of the event invitations (62 events, or 78% of the events) were sent through Facebook. Seven events were sent from particular event websites, and the rest were sent through other channels, such as instant massaging, BBS, or email.

Table 1 shows the interviewees' responses to event invitations they received. After excluding 12 cases in which the interviewee did not remember how s/he replied, of the remaining 67 events, 43 accepted, 19 ignored the invitation, one responded ambiguously (i.e., with uncertainty), and four declined. The 43 invitations that were accepted were reported by and distributed among 15 interviewees. Interestingly, of the 43 accepted event invitations, 17 (39.5% of accepted events) did not result in actual attendance.

The descriptive statistics show that the most popular form in which the interviewees received invitations was Facebook. We also confirmed the presence of a gap between online and offline participation. A high proportion of accepted events did not result in attendance, raising the question of why people committed to attending events online but skipped them offline.

**Roles of Interface Features**
Several salient themes emerged from the data. The first theme that emerged from the interviews is the role of interface features, which shaped the way people responded to the invitations. SNSs such Facebook and Google+ allow for explicit event creation and management, so people can initiate events, edit event details, and send invitations to invitees. SNSs also allow invitees to reply to invitations by clicking on buttons such as "Join," "Maybe," and "Decline" on Facebook or "Yes," "Maybe," and "No" on Google+.

The majority of the invitations that our interviewees reported were sent from Facebook, so our discussion concentrates on these. We describe how our interviewees perceived the interface features (e.g., buttons) of the invitations and the impacts of these features on their responses to the invitations. The interviewees were anonymized and are presented in the following format: "number-F/M-S/N" ("F": female, "M": male, "S": student, "N": non-student).

*Social Meanings of Reply Buttons*
Although Facebook provides a set of buttons to allow people to easily express their intention to participate, different people may interpret the meaning of these buttons differently. Twelve of our interviewees claimed that clicking the "Join" button on an event page is *not* a promise that one will actually attend the event.

*"This is not a contract that can force you to execute. I click 'Join' only to indicate the inclination. It implies that 'I most likely will attend' or 'I most likely won't go.'"* (1-M-S)

One useful function that Facebook provides is the ability to remind users of events that they have agreed to attend and notify them of updates to and the times of these events. Three interviewees appropriated this function for personal information management. They clicked on response buttons on Facebook event pages so that Facebook could help them remember the events rather than to indicate an intention to attend.

*"I clicked 'Join' because the event looked good. Basically, it was similar to writing a memo saying, 'I want to go to this event'...so if I can't attend at the time, I won't change my response on the page to 'Maybe' or 'Not going.'"* (30-F-S)

Similar to interviewee 1, interviewee 25 noted that the literal meaning of "Join" was overestimated and treated lightly, suggesting that people clicked on this button without a great deal of consideration. Decision making regarding event participation is taken less seriously on SNSs.

*"No one responds seriously. They take a look and think maybe they are available to go, and then click 'Join,' but forget about it within the next second."* (25-F-S)

In contrast to agreement, a refusal to participate is accompanied by considerable thought. Eleven interviewees thought seriously about the meaning of clicking the "Decline" button because they were concerned about the negative impact that it would have on the event or the host.

*"If the number of people who have declined is higher than the number of people who have joined, one thing I will consider is the host's feelings. The other thing is that it may cause others who see the page to wonder why they are going to an event no one wants to go to."* (10-M-N)

Furthermore, according to three interviewees' reports, "Decline" indicates not only an unwillingness to attend but also an opposition to the event. If they cannot make an event, they prefer to ignore the invitation rather than provide a definite negative response. Erroneous attribution regarding their stance on the event can thus be avoided.

*"If the event concerns some issues that I disapprove of, of course I will click 'Decline.' However, if I support its standpoints, I won't click 'Decline' whether I will attend or not."* (31-F-S)

*"If I click 'Decline,' it feels like it's because I think the one who invited me made some mistakes."* (07-F-S)

The meaning of the choice to "Join," "Decline," or not reply extends beyond what was suggested when our interviewees made the decision upon receiving an invitation. Although people differed in their evaluations and perceptions of the events they agreed to attend, they weighed a negative decision more heavily because rejection can have unintended social consequences. In the following, we discuss the interviewees' evaluations and perceptions used to reach their attendance decisions.

*Targeted versus Spam Communication*
Some interviewees had experiences of being both invitees and event hosts, so they were familiar with the procedure used to create events and the associated costs of initiating events on Facebook. Their perceptions of the cost of event management on different sites or through different media became a criterion for their evaluation of the importance of the events.

| Offline Behavior | Online Response | | | | |
|---|---|---|---|---|---|
| | Accept | Uncertain | Decline | Ignore | Total |
| Attendance | 26 (60.5%) | 0 (0%) | 0 (0%) | 2 (10.5%) | 18 |
| Absence | 17 (39.5%) | 1 (100%) | 4 (100%) | 17 (89.5%) | 39 |
| Total | 43 | 1 | 4 | 19 | 67 |

**Table 1. The number of specific events identified in the interviews by type of online response and offline behavior.**

*"Facebook provides a template for you to create events, and you just have to fill in the blanks. However, with websites, you have to establish it from scratch…If someone builds a website for the event, I think it must be serious, or the host won't be willing to build the website."* (09-F-S)

*"Currently, many people create events on Facebook, and if you want to go, click 'Join'; if not, then ignore it. I sense that many people do that, and the event can fall apart. However, if the event host invites me by phone, it indicates that s/he really takes the event and my attendance seriously; otherwise, s/he won't call everyone one by one."* (20-F-N)

Eleven interviewees noted that the low cost and the ease of inviting people on Facebook suggested that the host had not taken the invitation seriously, especially if the list of invitees is large and all-inclusive. In such cases, the response was weighed as much as the effort the event hosts spent on the events; the invitees tended to treat such events lightly and ignored the invitations.

*"When you use Facebook, you can invite all your friends on your list with one click. It can be done as easily as spam. However, with e-mail, you have to screen who may be more interested in the event and even write down some words for the invitation."* (31-F-S)

*"If it's instant messaging, that's a personal invitation. That is more sincere and is not a random attempt."* (19-M-N)

If the invitations are not sufficiently targeted, people's intentions to reply or attend may be low. Interviewees compared invitations sent through different media. The data show that if a public medium is used and if it is easy to invite participants, the invitation appears less formal and less personal.

*"I think email can be replied to personally, and it's private. Facebook seems still semi-public even when we use group chat. I think a Facebook invitation is more casual in that the event hosts invite participants by clicking on them without concern. However, email may require event hosts to search for the list. Even if they just select all the people on the list, they should have contacted you before, and that makes me feel…it's more close to me."* (28-F-S)

If an invitation was more personal, the interviewees took it more seriously and tended to provide an unambiguous answer instead of ignoring the invitation. In other words, the personalization of the invitation has a positive relationship with perceived accountability [32]. In our data, the level of personalization was evaluated based on the media used to send the invitation. Our interviewees applied different strategies when replying to invitations sent through different media.

*"I always indicate clearly whether I will go or not in response to invitations sent through instant messaging, as it feels closer to being invited on a one-on-one basis."* (30-F-S)

To our interviewees, the perception of the directness of the invitation/communication was vital to the decision regarding attendance and commitment to the decision. The undirected approach of invitation dissemination, although efficient on the hosts' end, was taken less seriously by the invitees.

**Public Events with a Mass Audience**
In addition to the perception of whether an invitation is communicated in an undirected or directed way, the nature of the event, such as whether it is a public or a private activity, elicits different reply strategies from users. Public events on Facebook allow access to everyone (based on privacy settings) and may feature an invitee list or participant list with hundreds of people. We found that the public nature of the event and the visibility of the participant and invitee lists may affect people's intentions to respond to such an invitation and their offline participation.

*Public Image Maintenance and Conflict Avoidance*
On Facebook, if individuals "Join" an event set as "Public," information regarding their attendance is shown on their personal walls by default. When managing invitations, our interviewees' changed their behaviors in response to such a public display to maintain their public images. Sometimes they even use such an invitation as an opportunity to reveal their tastes or interests to others. This online representation is expected to facilitate conversation with an interested audience. The online display of event participation is also considered a less socially intrusive method of self-promotion.

*"If there is an Audi or Mercedes-Benz exhibition, I will click 'Join' to let others know, 'Wow, it is so cool that you go to this event,' and it opens conversation…but if you have no interest in this, and I keep telling you directly that I went to a Mercedes-Benz exhibition today, or posting it on others' walls, people will think I am nuts."* (10-M-N)

However, if the event was related to controversial issues (e.g., anti-nuclear power, anti-media monopoly in Taiwan) or was in conflict with their online personal identity, the interviewees tended to be reserved in terms of expressing their intention to join.

*"Everyone has his/her own standpoint, and I don't want to be labeled. To some degree, I am afraid that my friends will think I am a certain kind of person because I join some events. Therefore, I will avoid clicking 'Join' for an event that states certain positions."* (02-M-S)

Event attendance becomes part of one's online representation. To display an ideal image, nine interviewees reported that they attempted to show the best part of themselves and avoided possible conflicts. One way to ease the tension resulting from this conflict is to remain within the mainstream. Due to the bandwagon effect [24], the majority opinion prompts people to side with the majority view. This effect relieves people of the pressure of taking an explicit stance or incurring conflict.

*"From the side of public opinion, if everyone supports this event, of course you will click 'Join.' It means that I have the same opinion as the mainstream."* (01-M-S)

The discrepancy between online and offline representation sometimes produces an inconsistency between online and offline participation. To protect their privacy, individuals may not click on "Join" online but may still attend the event offline. One of the interviewees described his experience regarding an invitation to a rock festival. He went to the event with his friends without accepting the invitation online because he did not want others to know what he was doing (24-M-S).

The negotiation between online and offline images is especially apparent in situations in which people use Facebook to maintain offline relationships [11, 22]. People feel tension when dealing offline with the consequences of their online behaviors.

*"Because my Facebook friend list consists of my teachers, relatives, and elders, they are going to tell me that they see that I clicked on the anti-nuke demonstration or something, and ask questions like, 'How can you students be so available as to join such events?'"* (09-F-S)

In alignment with the notion of context collapse [39], people cannot easily adapt their self-presentation to different audiences and contexts with a single, shared information space (e.g., a Facebook wall). The multiplex networks on SNSs make it difficult for the interviewees to manage a consistent online/offline presentation. As a result, our interviewees employed other strategies, such as following the majority view or hiding their real thoughts, to respond to public event invitations to maintain their ideal self-representation on the front stage.

*Social Loafing*
Studies on social loafing posit that individuals reduce their group work efforts when the number of people in the group grows [13]. On a Facebook event page, the names and number of participants and invitees are disclosed. Ten interviewees mentioned that they were aware of the number of other invitees and attendees. Hence, they were less concerned about their response because they believed that no one would notice their absence and thus did not feel pressured to attend the event. The display of this information appears to result in social loafing.

*"When there are only a few people, maybe just five, being invited, every decision of each individual is obvious. However, it doesn't really matter if someone clicks 'Join' or not in the event with 1,000 people invited."* (01-M-S)

A large number of attendees may lead to social loafing in a form similar to slacktivism when the event concerns political movements. Social loafing may explain the difficulty of de-virtualizing such online events. People depend on others' actions without taking theirs seriously or think that their actions do not count as much when the denominator is large.

*"For the recent anti-nuke demonstration, I clicked 'Join' on the event page, but just for moral support. That event did not really need my actual attendance."* (22-M-S)

In other words, a perception of a large number of both *invitees* and *attendees* may trigger social loafing during the decision and action stages.

**Within-Group Private Events**
Based on the descriptions given by our interviewees, private events restrict information access to those who are invited. Usually, private events are held within private social groups (e.g., high school classmates, family) containing individuals who know one another and have had prior offline social interactions. A sense of social obligation among individuals may create pressure to honor their commitment to attendance. In addition, a prior relationship implies that the host has multiple ways of contacting the participants, such as phone calls, emails, or IM. Multiple ways of staying in touch may also contribute to increased levels of offline participation.

*Social Obligation*
For private events, both the host and the participants are usually people who meet each other offline and have an expectation of persistent future interactions, which results in a sense of social obligation and prompts more definite responses. Twelve interviewees reported that, unlike public events for which they tended to ignore invitations, they gave precise responses for private events.

*"For a private event, I will indicate clearly whether I will join or not. However, for those public events, I do nothing if I won't engage."* (02-M-S)

In addition to the decision stage, social obligation plays an important role at the action stage for private events. Most of our interviewees reported that they would not be absent from such events because they were worried about being blamed by the members of the social group. In other words, the expected interaction and identifiability in the group caused our interviewees to self-censor when making a decision to avoid possible social sanctions.

*"In such a private event, honestly, everyone knows each other. You can't act like you did for the public events— clicking 'Join' but being a no-show. That will get you in trouble."* (23-M-S)

*"Like a class reunion, the participants are all my friends, and I think the host relies on the Facebook event page to do the head counts. No-shows will cause them trouble."* (30-F-S)

In comparison with public events, private events cause people to care about their decision, commitment, and action, either for their sake or for the hosts' sake.

*Confirmation through Different Media*
Prior relationships usually indicate that people have multiple ways of connecting to each other. Ten interviewees talked about the experience of making use of such connections to ensure participation. For example, the host can make a phone call to everyone on the "Going" list for double confirmation (if one clicks on "Join" on the event page, his/her name and photo show up on the "Going" list).

*"My friends, especially very close ones, they create the Facebook events just for fun. A phone call is coming up anyway, for they think I may miss the information."* (16-M-S)

In addition to Facebook events, other online communication tools are used to raise awareness of the event or to confirm attendance.

*"Facebook event invitations are not completely reliable. I most likely send an instant message through both Facebook and Windows Live Messenger, even send a Line message by cellphone."* (18-M-N)

*"Before the event, we will open a chat room in Facebook message, and it will present who has seen the reminder message. We use it confirm that everyone has gotten the information."* (07-F-S)

The event planning process does not always move from online to offline. Some interviewees noted that the reverse route produced even stronger confirmation of attendance. People proposed an event and counted who would join in advance when they met offline. As the meet-up time approached, they created a Facebook event with more details regarding time and location and other information to raise awareness and receive final confirmation.

*"Facebook is like an assisting tool… I will make sure my event guarantees a certain level of attendance before I create it."* (06-M-S)

*"For all the private events with usually a small number of people, first, we ascertain we are going out together; then, we create Facebook events for a real confirmation of attendance."* (07-F-S)

Our data show that using different communication tools to convey a message or combining online and offline communication are ways of raising people's awareness of and commitment to an event and increasing the likelihood of response and attendance.

**Social Ties with Others**
Regardless of whether an event is public or private, people's decisions regarding the event invitation appear to differ depending on whether it is received from a friend or an acquaintance.

*"It depends on who invited you. If he/she is not so important, I will most likely click 'Maybe' or just ignore the invitation. If he/she is important or is my close friend, I am certain to reply. Moreover, for events for which I'm not sure whether I can go, I will click 'Join' and leave a comment like, 'I'm not sure if I am available that day, but I will try my best to attend' on the event page."* (21-F-N)

Strong social ties entail mutual trust, which influences invitees' evaluations of the information (e.g., the number of participants) on the event page. However, even if the event is public, being invited through strong ties increases the possibility of eventual attendance.

*"Today, I create an event and tag others' names, hoping they reply to the invitation. For people tagged, they can't miss my message. I just treat those who do not respond as if they are not going to join us to do the head counts. The thing is, it only works when you know all your invitees. If there is a stranger, he is the weakest link."* (23-M-S)

The interviews also indicate that strong ties with the attendees are important to the host as well because these relationships help to guarantee attention to and attendance at events.

*Motivating Offline Participation*
Public events can lead to social loafing. However, strong social ties can be a solution to this problem. Six interviewees described how seeing other friends on the "Going" list could improve people's intentions to join a public event and their subsequent joining behaviors.

*"If my close friends are on the list, I will send instant messages to ask them. If they are certain to go, okay, then I will go, too."* (09-F-S)

*"When I first see the (public) event, I see if my friends are going…I will ask them how the event is. If it is fun, then I will join."* (23-M-S)

Because of the information that is publicly displayed on Facebook, people who know each other can form a small group and attend a public event. Offline social relationships facilitate the de-virtualization of online participation into offline participation.

*"When there are thousands of people clicking 'Join', to be honest, your attendance does not matter, unless there is someone who draws you in. For example, you see your friend is on the list, and that friend you haven't met for a long time…Therefore, it's often the case that a couple of friends attend the social movement together. This helps maintain the offline attendance level."* (11-M-N)

*Difficulty of Declining*

Social obligation derived from relationships characterized by strong ties not only encourage people to participate offline but also make it difficult for interviewees to decline an invitation because "Decline" may imply disapproval of the host or the event, as previously mentioned.

*"I found that if I click 'Decline', the event holders will be disappointed because they assume that I do not want to go rather than that I have no time. Therefore, I click neither 'Join' nor 'Decline' in the end."* (4-F-S)

In a way, social relationships between the host and the invitees sanction the invitees' decision. When responding, people prioritize the relationship over their interest in the event or their availability.

*"For acquaintances, I click 'Decline'. For close friends, I click 'Maybe' or 'Join', but if I won't go, I make a phone call to explain why I can't go."* (16-M-S)

Seven interviewees viewed the choice of "Decline" as an impolite and thoughtless behavior, especially when the event was initiated by people with whom they had strong connections. Therefore, they clicked "Join" out of courtesy, not as an indication of commitment or attendance.

*"Our department created an event last afternoon, and I clicked 'Join' merely out of courtesy."* (02-M-S)

An invitation through strong ties is a double-edged sword; it helps to ensure attendance but introduces uncertainty to offline attendance. Seven interviewees indicated a sense of social obligation to accept invitations although they were sure that they would not attend, allowing them to save face for the host and avoid social sanctions.

*"Information on Facebook is visible to everyone, and I think it could be a problem for that...I saw my friends all clicked 'Join', and I found it embarrassing if I didn't. Then, I clicked it, but I wouldn't go."* (12-M-S)

*"...if he asked me to come, of course I said, 'Yes, yes, I will'. You have to support him, right? So I click 'Join', but I ultimately didn't attend."* (06-M-S)

**DISCUSSION**

Based on the events identified in the interview transcripts, different types of online responses and actual attendance show that a gap exists between online and offline participation ("the gap," hereafter). Other recent works have identified similar results [12, 25, 35, 41]. In socio-psychological research, the association between attitude and behavior is unclear [1]. Attitudes displayed are not necessarily consistent with behaviors. Although technologies such as SNSs and CMC provide convenient tools with which people can express their support and commitment, what people say (online participation) is not necessarily equal to what they do (offline participation).

The current work explored why this gap occurs and clarifies how and why the de-virtualization of social events at the individual level may be difficult. We found that perceptions regarding interface features and communication platforms (e.g., the meanings of reply buttons on Facebook), the social nature of the event (e.g., public versus private events), and social ties (e.g., relationships to the event host and other invitees/attendees) are important factors.

The first factor in the gap is related to how people perceive the meanings of the interface features of the communication platforms used to communicate events. Surprisingly, people can attach non-literal, social meaning to response buttons (accept, decline, maybe). For example, clicking "Accept" on Facebook becomes a way to express attendance uncertainty to save face for both the host and the invitee. The selection of "Decline" expresses one's opposition to the event. Such appropriation can lead to the miscommunication of intention, especially when the invitee and the host do not share the same interpretation frameworks. It appears that people may form implicit norms regarding the social meanings of these buttons. However, because this sort of social interpretation is implicit, inconsistency in interpretations may create false expectations and inaccurate estimations of the number of actual attendees at offline social events. Current interfaces do not support the nuanced appropriation and interpretation of the choice buttons. A lack of design subtlety gives users less flexibility to express their genuine intentions online that accompany their offline behavior.

It is also interesting that the interviewees were aware of the costs associated with online event communication. The literature focuses primarily on the costs associated with offline participation and considers online participation a low-cost and low-risk activity [30]. The cost-efficiency of online participation may reduce people's interest in participating in offline activities [25]. However, event communication and arrangement involve effort from the host in addition to the invitees. With CMC and SNSs, both the cost of participation and the cost of event planning decrease. The host's lack of effort in initiating event invitations on SNSs appears to lessen invitees' evaluations of the event.

For event invitations targeted at a large audience, the invitation is easily distributed with the aid of features provided by SNSs. For example, on Facebook, little effort is required to recruit participants using features provided on the event page. When the cost of recruitment is low, invitees are unlikely to reciprocate through offline attendance; they may return a mouse click, at most.

Second, we found that people use different response strategies for public and private events. The social nature of an event has an impact on people's participation because it triggers perceptions of being part of a large crowd or a small group and indicates how identifiable the individual is in the context of the social event.

With online event invitations, information that is displayed regarding how many people plan to attend the event and these people's identities is an important cue that influences invitees' behaviors. Our interviewees reported a weaker motivation to join a public event when they knew that many people were attending. Because the cost of online participation on SNSs is low, a new form of social loafing behavior emerges in the form of an online expression of support (i.e., clicking "Join" on the Facebook event page but failing to actually attend offline). Slacktivism with regard to political events may occur for this reason. Conversely, an underestimation of the number of participants may trigger participation [27]. The perceived size of participation, either overestimation or underestimation, appears to influence individual action.

Private event invitations tend to de-virtualize social events relatively better, channeling people's online commitment to offline attendance. Consistent with work on the mobilization of online communicators [3, 32], tie strength appears to correlate with whether people will join an event. People may be more willing to attend an event when they discern that their close friends also plan to do so, suggesting that existing connections and community bonds play a role. At the same time, because people also use Facebook to sustain their relationships with people they have met offline [22], an influence of strong ties is present online. One possibility is that information disseminated via these strong ties may easily dominate one's online information processing, resulting in limited attention to weak tie-mediated event information. The mobilizing effect of event invitations mediated by weak ties may thus be further attenuated. Although we believed that interaction on SNSs may facilitate socialization and may compensate for the weakness of weak ties in the de-virtualization of social events, we did not observe cases supporting this assumption. Further design efforts may be required to support this scenario.

Relationship concerns can be a double-edged sword. Our data reveal that strong ties may promote offline participation through perceived social obligation, but it may also become difficult for people to decline events that they would not or should not attend. Although extensive research has examined the relationship between social ties and participation [3, 8, 15, 16, 33, 41], the negative side effects have received limited attention.

Impression management is another concern on SNSs because participation information is displayed to other invitees or on one's personal network. Previous studies have shown that individuals leverage SNS profiles and online group affiliation as resources to shape their online self-presentations [9, 12, 15]. Our study sheds light on how concerns about impression management interact with the content and context of events to affect people's event participation. For example, people choose to participate in or avoid certain events online or offline to maintain their image.

People can develop sophisticated strategies for event participation to meet their need for impression management, such as offline attendance at an event without an online expression of support to avoid an explicit affiliation with any stance. People also have a need to differentiate their online image for different groups of SNS contacts (e.g., family, colleagues, close friends), such as by displaying event participation information to some parties but not others.

**Comparison between Invitations via Different Media**

In our results, the media used and the perceived sincerity of event hosts constituted one reason for the gap between online and offline participation. Interviewees identified the differences among invitations sent via different media, such as Facebook events, instant messaging, phone, email, and face-to-face communication. Based on the evaluation of the media used, they determined whether the invitations were important enough to merit a reply. In addition, they made attendance decisions according to the perceived costs of event arrangement. Our data show that when people receive invitations through relatively personal and private media, such as face-to-face interactions, phone, or even instant messaging, they feel that the event hosts care significantly more about their attendance than they would if they used a more social and public medium, such as a Facebook invitation. They know that the event hosts have made a special effort to invite them instead of randomly selecting individuals from long lists. At the same time, social loafing is less likely to occur with offline acceptance or online instant messages because people cannot determine the number of participants, which is disclosed on SNSs, or the hosts are individuals with whom they have consistent interactions. Therefore, they make their decision more carefully and are more likely to keep their promise of offline participation. The attitude toward the event indicated by their online participation corresponds more closely to the targeted offline participation behavior [1].

Face management in SNSs corresponds closely to Brown and Levinson's study, in which people developed politeness strategies to avoid making others feel uncomfortable and to maintain others' self-esteem [6]. We found that rather than directly declining invitations in SNSs, people often employ off-the-record strategies of refusal by expressing their intentions indirectly offline or via other media (e.g., instant messaging). As one interviewee explained, she turned down a wedding invitation over the phone. She made an excuse such as, "*I am not sure if I will be available that day. Let me check my schedule*" and then called back later to apologize because she could not attend the wedding (26-F-N). This sort of complicated decision-making and communication process is reduced to the click of a button on SNSs. The interface design may make people feel that it is difficult to decline event calls; thus, they click "Join" to maintain politeness. Current SNS designs may not be able to support the intricate social protocols behind declining invitations that people practice offline.

## DESIGN PROPOSALS

Based on the results, we propose a number of initial ideas for supporting people in extending their online participation to offline attendance and for better de-virtualization of social events. We believe that the design space constructed on the obtained understanding is significant and sophisticated. Therefore, our proposals are intended to demonstrate rather than to limit ways to design and redesign tools for event management and communication.

First, to increase the effectiveness of online event invitations and to encourage offline participation, it may be necessary to explicitly present the host's effort and the value of participation in the event to the invitees. For example, the SNS's interface for event creation may request that event creators provide more detailed and personal invitation messages, and the human effort of personalization must be visible to the invitees.

Another proposal involves making people feel more comfortable declining an event invitation so that invitees' online responses are less ambiguous and more reliable for event management. One respondent indicated a possible strategy: *"I click 'Join' first and then post a message to apologize that I won't attend"* (2-M-S). Instead of relying on a single button (Accept, Decline, etc.) to express a decision, clearly distinguished buttons can indicate support of the event (such as a thumbs-up) and the intention to attend. Alternatively, both the invitees and the event host could be allowed to express their intentions flexibly with open-ended, short messages. This design could also use text processing and machine learning techniques to automatically estimate the likelihood of actual attendance. In this case, machine processing could be used to aid people in maintaining the desired ambiguity.

Because people tend to loaf more in larger groups [20] and are more willing to attend private events than public ones, it may be effective to manipulate perceived group size and privacy through design. One possible design involves the transformation of public events into more private ones, such as by partitioning a loosely connected event group into multiple subgroups in which social connections among group members become relatively stronger and more private. In this way, group size decreases and the perception of privacy increases.

Because information on event participation on SNSs can influence how one is viewed and valued by other people, it may be helpful to consider the benefits and costs of displaying information on online participation (e.g., attending an event) and offline participation (e.g., a status or photo indicating physical presence) as well as the (in)consistency between the two.

## LIMITATIONS

This study has two main limitations. First, because this study is an interview study, data collection is inherently connected to the properties and constraints of self-reporting. The recollection of information related to event participation through memory recall may not be completely reliable, and interesting phenomena may be missing from our interviews. However, the themes that emerged from the current transcripts appear to make a great deal of sense and have triggered useful insights. Future investigation will benefit from the use of other methods in the triangulation of our results.

Second, because our interviewees are Taiwanese living in Taiwan, the results reported may be culture-specific. Although limited clues allow for the determination of whether the perceptions, thoughts, and behaviors related to event participation identified in this work are specific to the domestic culture or are universal, it is reasonable to suggest that culture may play a role in group-related processes such as social loafing and obligation. The cross-cultural aspect is especially worth further investigation.

## CONCLUSION

This paper presents an interview study intended to increase understanding of people's experience and practice handling of distributed event invitations. From these interviews, we gained a richer and deeper understanding of how people handle online event invitations, the shape of online and offline participation behaviors, and the gap between the two. We developed the notion of de-virtualization to describe social events that are developed online but that aim to provoke participation offline, and we examined why this may be a challenging task and how we can address this challenge through design.

The results reveal reasons for participation and non-participation both online and offline. The wide dissemination of e-invitations does not necessarily guarantee high participation. This study contributes to the literature by showing that social perceptions of the nature of an event (e.g., public or private), relationships to others (e.g., the strength of connections to other invitees), and the medium used to communicate event information (e.g., targeted invitation via email or spam communication via a Facebook event page) are possible key factors. Future designs may leverage these understandings to better bridge people's online and offline participation in events that require physical presence and real-world attendance.